\documentclass[twocolumn]{aastex62}

\graphicspath{{./}{figures/}}

\submitjournal{ApJL}

\shorttitle{HD 98800 polar disk}
\shortauthors{Franchini et al.}

\begin{document}

\title{Circumbinary disk inner radius as a diagnostic for disk--binary misalignment}

\correspondingauthor{Alessia Franchini}
\email{alessia.franchini@unlv.edu}

\author{Alessia Franchini}
\affil{Department of Physics and Astronomy, University of Nevada
4505 South Maryland Parkway, Las Vegas, NV 89154, USA}

\author{Stephen H. Lubow}
\affiliation{Space Telescope Science Institute
3700 San Martin Drive, Baltimore, MD 21218, USA}

\author{Rebecca G. Martin}
\affiliation{Department of Physics and Astronomy, University of Nevada
4505 South Maryland Parkway, Las Vegas, NV 89154, USA}



\begin{abstract}
We investigate the misalignment of the circumbinary disk around the binary HD 98800 BaBb with eccentricity $e\simeq 0.8$. \cite{Kennedy2019} observed the disk to be either at an inclination of $48^{\circ}$ or polar aligned to the binary orbital plane. Their simulations showed that alignment from $48^\circ$ to a polar configuration can take place on a shorter timescale than the age of this system.  We perform hydrodynamical numerical simulations in order to estimate the cavity size carved by the eccentric binary for different disk inclinations as an independent check of polar alignment.  Resonance theory suggests that torques on the inner parts of a polar
 disk around such a highly eccentric binary are much weaker than in the coplanar case, indicating a significantly smaller central cavity than in the coplanar case.
We show that the inferred inner radius (from carbon monoxide measurements) of the accretion disk around BaBb can exclude the possibility of it being mildly inclined with respect to the binary orbital plane and therefore confirm the polar configuration.
This study constitutes an important diagnostic for misaligned circumbinary disks, since it potentially allows us to infer the disk inclination from observed gas disk inner radii.

\end{abstract}

\keywords{accretion, accretion disks --- 
binaries:general --- hydrodynamics --- planets and satellites:formation}


\section{Introduction} 
\label{sec:intro}

The majority of stars form in binary or multiple systems \citep{Ghez1993} and circumbinary disks, as well as circumstellar disks, are likely to form. 
Stars do form in a chaotic environment \citep{McKee2007} and therefore we expect misaligned circumbinary disks to be quite common rather than rare \citep{Offner2010,Bate2012,Bate2018}.

The evolution of circumbinary disk orientations have been extensively investigated in previous works. 
If the binary orbit is circular, an initially misaligned circumbinary disk precesses about the binary angular momentum vector and eventually aligns with the binary orbital plane \citep{Papaloizou1995,Lubow2000,Nixon2011,Foucart2014}.
If the binary orbit is eccentric and the disk misalignment is above a critical inclination, the circumbinary disk precesses around the eccentricity vector of the binary and aligns its angular momentum to it  \citep{Aly2015,Martin2017,Lubow2018,Zanazzi2018,Martin2018}. This case is referred to as polar alignment.

The torque exerted on an aligned (prograde and coplanar) circumbinary disk by the binary opens a central gap. The gap has been directly observed in some cases, such as in GG Tau
\citep[e.g.,][]{Dutrey2016}. The gap is opened by the effects
of torques due to Lindblad resonances in the disk. The gap size is a function of the binary separation, eccentricity, and disk viscosity \citep{Artymowicz1994}.

If the disk is not aligned with the binary, there is a reduction
in the torque and consequently a reduction in the central gap size.
This is expected for two reasons. First, misaligned disks feel a weaker Lindblad torque from the binary \citep{Lubow2015,Nixon2015,Miranda2015}.
For significant binary-disk misalignments, this effect occurs because the disk material at some radius can be on average 
located farther away from the binary than would occur for a coplanar disk and in addition the relative motion of the binary and disk increases. The latter effect is especially important for retrograde disks \citep[e.g.,][]{Nixon2015}.
Second, the Lindblad torque for a highly misaligned disk
decreases with binary eccentricity, if the eccentricity is high \citep{Lubow2018}. This latter effect can be understood in the extreme case of a polar disk with a central binary on an orbit with eccentricity of unity. In that case, at each instant in time the binary potential is axisymmetric in the plane of the disk and therefore the Lindblad torque on the disk is zero. 
We find that for tidal azimuthal wavenumbers $m \ge 2$, the Lindblad
torque on a polar disk approaches zero as $(1-e)^m$ for binary eccentricity $e$ close to unity.
In contrast, an aligned binary-disk system involving a highly eccentric binary produces
a strongly nonaxisymmetric potential in the plane of the disk, generally resulting
in a strong Lindblad torque on the disk.
Since polar circumbinary disks are more likely associated with high eccentricity binaries, this effect should be quite important in reducing its central gap size relative to the gap in a coplanar disk.

Many misaligned circumbinary disks have been observed so far at different stages in the evolution of binary systems.
The extensively studied pre-main sequence binary KH 15D hosts an inclined and precessing circumbinary disk \citep{Winn2004,Chiang2004,Capelo2012,Smallwood2019}. The circumbinary disk that orbits around the binary IRS 43 is highly misaligned with respect to the binary orbital plane \citep{Brinch2016}.
The most extreme misaligned circumbinary debris disk observed so far is the debris disk around the eccentric binary 99 Herculis that has been inferred to be polar (i.e. with its angular momentum along the eccentricity vector of the binary) \citep{Kennedy2012}.

Very recently the ALMA (Atacama Large Millimiter/Submillimiter Array) telescope allowed a more precise characterization of the highly misaligned circumbinary disk around one of the two components of the quadruple system HD 98800 \citep{Kennedy2019}.
This system is composed of two pairs of binaries ('A' and 'B', or 'AaAb' and 'BaBb') orbiting each other with semi-major axis $54$ au, eccentricity $e_{\rm AB}=0.52\pm0.01$ and period of $246\pm 5$ years.

The binary BaBb has a very well characterized orbit with semi-major axis $1$ au and eccentricity $e=0.785\pm 0.005$. The masses of the two stars are $M_{\rm Ba}=0.699\,M_{\odot}$ and $M_{\rm Bb}=0.582\,M_{\odot}$  \citep{Boden2005,Verrier2008}.
HD 98800 BaBb has been known to host a bright circumbinary disk \citep{Walker1988} that was initially inferred to be coplanar with the binary orbital plane \citep{Andrews2010}. However, higher resolution images obtained with ALMA showed that the disk orientation is quite different \citep{Kennedy2019}.

From the modelling of the dust and carbon monoxide measurements, \cite{Kennedy2019} inferred a disk radial extent from $2.5\pm 0.02$ au to $4.6\pm 0.01$ au in dust and from $1.6\pm0.3$ au to $6.4\pm0.5$ au in gas.
Both components have been found to be consistent with having the same orientation. The disk inclination relative to the binary orbital plane has been inferred to be either $48^{\circ}$ or close to $90^{\circ}$.
\cite{Kennedy2019} further investigated this system by performing numerical Smoothed Particle Hydrodynamics (SPH) simulations. They modelled two disk orientations, $48^{\circ}$ and $90^{\circ}$ (polar), and concluded that the polar configuration is the more likely based on the argument  that the time required for a disk with initial misalignment of $48^{\circ}$  
to evolve to $90^\circ$ (polar) 
is very short compared to the age of the stars. 

As also pointed out in \cite{Kennedy2019}, there is some uncertainty in the evolution time due to unknown properties of the disk, such as its turbulence parameter $\alpha$. 
For sufficiently small $\alpha  \stackrel{<}{\sim} 10^{-5}$, linear models
suggest that the tilt evolution timescale could be comparable to the age of the system of 10 Myr \citep{Lubow2018, Zanazzi2018}. 
While we are not suggesting these conditions are likely in HD 98800,
they may be possible. 
We are therefore motivated to provide an independent check on the inclination of this disk. Furthermore, such an independent check may be useful in other systems where the parameters are not as well observationally determined. 

In this Letter we show that a circumbinary disk orbiting the binary BaBb can radially extend in to the observed inner radius only if the disk is in a polar configuration.
We therefore confirm the conclusion drawn in \cite{Kennedy2019} that the disk is in a polar configuration.
This analysis constitutes an important diagnostic for this type of observations and allows us to exclude the possibility of the disk being inclined by $48^{\circ}$ with respect to the binary orbital plane.

\begin{figure*}
\gridline{\fig{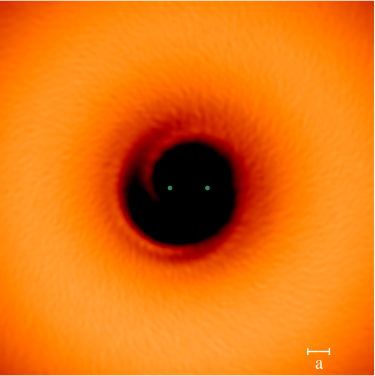}{0.3\textwidth}{}
          \fig{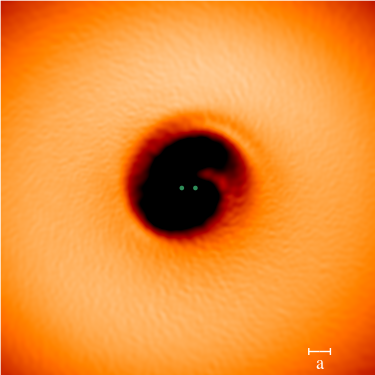}{0.3\textwidth}{}
          \fig{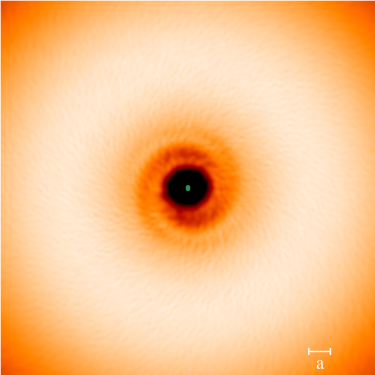}{0.3\textwidth}{}
          }
\caption{\label{fig:phantom} Circumbinary accretion disks around the binary BaBb viewed face on.  The disk is shown at time of $740$ binary orbits with inclination: $i=0^{\circ}$ (left panel), $i=48^{\circ}$ (middle panel) and $i=90^{\circ}$ (right panel). The binary components are shown by the two green circles. In the polar aligned disk on the right, one star is behind the other.  }
\end{figure*}

\begin{figure}
\includegraphics[width=\columnwidth]{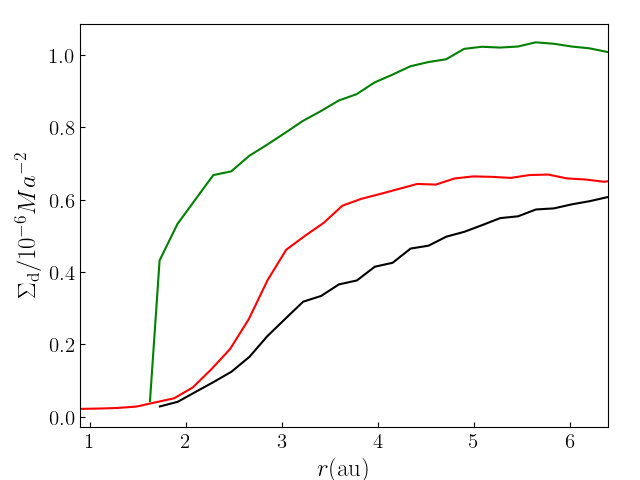}
\caption{\label{fig:surfacedens} Surface density profile of the circumbinary disk after $740$ binary orbits for three different inclinations: $i=0^{\circ}$ (black), $i=48^{\circ}$ (red) and $i=90^{\circ}$ (green). The surface density $\Sigma$ is normalized by  $10^{-6} M/a^{2}$, the binary mass devided by the square of its separation.}
\end{figure}

\section{Hydrodynamical simulations} 
\label{sec:sim}

We ran SPH numerical simulations using the code {\sc phantom} \citep{Lodato2010,Price2010,Price2012,Price2017}.  Misaligned disks in binary systems have been extensively studied with {\sc phantom} (e.g. \cite{Nixon2012,Nixon2015,Franchini2019}). We perform SPH simulations using $N=5 \times 10^5$ particles. 
The resolution of the simulation depends on $N$, the viscosity parameter $\alpha$ and the disk scale height $H$. 
The \cite{SS1973} viscosity parameter is modelled by adapting artificial viscosity according to the approach of \cite{Lodato2010}. 

We set up a circumbinary disk orbiting around the binary BaBb.
The initial surface density profile is set as $\Sigma\propto (R/R_{\rm in})^{-3/2}$ between $R_{\rm in}=1.6$ au and $R_{\rm out}=6.4$ au and the sound speed distribution is given by $c_{\rm s}\propto (R/R_{\rm in})^q$ with $q=3/4$ in order to ensure that the disk is uniformly resolved \citep{Lodato2007}.
The initial mass of the gaseous accretion disk is set as $M_{\rm d}=0.001\,M_{\odot}$. The small disk mass does not significantly affect the evolution of the binary orbit \citep[e.g.,][]{Martin2019}. The aspect ratio at the inner edge is $H/R(R_{\rm in})=0.1$ and the viscosity is set as $\alpha=0.1$. Since these parameters are both upper limits to values expected for protoplanetary disks, we are considering the most viscous case possible. The cavity size is determined by the balance between the viscous torque with the binary torque and thus  these parameters lead to the smallest possible cavity size. 

We measure the size of the cavity carved by the binary in disks at different inclinations $i=0^{\circ},\,48^{\circ},\,90^{\circ}$. The disk has to reach a steady state in the inner parts of the disk where the tidal torque balances the viscous torque in the disk.  The inclination of the coplanar and polar aligned disks do not evolve and so finding  a steady state is straightforward. However, in the misaligned case, the inclination of the disk changes in time. 

A low mass circumbinary disk 
that is initially inclined by $48^{\circ}$  might evolve towards polar alignment, depending also on its initial longitude of ascending node,  and by this time the cavity is the same size as the disk that starts at $90^{\circ}$.
Therefore, we start a simulation with the circumbinary disk inclined by $20^{\circ}$. The disk will eventually evolve to coplanar alignment, but because the binary is eccentric, the disk undergoes tilt oscillations \citep[see for example Figures 6 and 12 in][]{Smallwood2019}. The disk becomes warped but reaches a density averaged inclination of $48^{\circ}$ at times of roughly $100$ and $740$ binary orbits. The disk surface density in the inner regions has reached a steady state by 740 binary orbits and so we choose to show all three simulations at this time.

Figure \ref{fig:phantom} shows the disk column density for the coplanar (left panel), $48^{\circ}$ (middle panel) and polar (right panel) configuration respectively after $740$ binary orbits. Each disk is viewed face on.
The binary components are identified by the green circles and the gas colour scale corresponds to roughly two orders of magnitude higher density in the yellow regions compared to the red ones.
We can clearly see that the cavity size decreases with misalignment angle. This result agrees with our expectations based on resonance theory, as discussed in the Introduction. 

Figure \ref{fig:surfacedens} shows the circumbinary disk surface density profile after 740 binary orbits for the three different simulations.  The black line corresponds to the coplanar disk, the green and red lines identify the disk at $i=48^{\circ}$ and $i=90^{\circ}$ respectively. 
The disc masses differ somewhat in the three cases likely, at least in part, because  of the initially small inner disc edge that causes more mass to be initially accreted onto the binary in the more aligned cases which have stronger torques than the polar case. 
We find that the disk is able to reach the observed inner edge of $ R_{\rm in} \approx 1.6$ au only in the polar case, while it remains truncated farther out (roughly $2.5$ au) in the coplanar and $48^{\circ}$ case. 



\section{Discussion and conclusion} 
\label{sec:disc}

The theory of disk resonances suggests that polar disks around highly eccentric binaries should contain considerably smaller central gaps than 
in the case of coplanar systems, as we find in our simulations.
We have  investigated the evolution of the surface density of the inner parts of a circumbinary disk around the binary star BaBb in the quadruple system HD 98800 for different disk inclinations. 
We have shown that the inner radius inferred from carbon monoxide measurements suggests that the only configuration possible is a polar aligned accretion disk.
According to the observations there are two possibilities: a disk misaligned to the binary orbital plane by  $48^{\circ}$ or a polar aligned disk \citep{Kennedy2019}.
However, the size of cavity is still larger in the $48^{\circ}$ case, therefore this type of configuration can be ruled out by the observations.
Furthermore, for HD 98800, the disk that orbits the highly eccentric binary and is initially inclined by $48^{\circ}$  can evolve towards polar alignment  on a  timescale that is short compared  to the age of the stars \citep{Kennedy2019}.  
Note that we considered the most viscous case possible and this gives the smallest possible cavity size.  More accurately determined CO density profiles would help further constrain
our models.

We showed, using hydrodynamical simulations, that the radial extent of the gas disk detected around the BaBb binary in HD 98800 is consistent with it being in a polar configuration and thus confirms the orientation found by  \cite{Kennedy2019}.
Although our analysis was applied to a particular system, it is important in general, since it provides a means of diagnostic for other circumbinary disks with poorly constrained inclination,  especially for highly eccentric orbit binaries. 
For disks whose gaseous configuration can be estimated from carbon monoxide measurements the analysis presented can be used as a tool to infer the disk inclination with respect to the binary orbital plane.

\section*{Acknowledgments}

We thank Daniel Price for providing the {\sc phantom} code for SPH simulations and acknowledge the use of {\sc splash} \citep{Price2007} for the rendering of the figures. We acknowledge support from NASA through grants NNX17AB96G and 80NSSC19K0443.  Computer support was provided by UNLV's National Supercomputing Center.

\end{document}